\documentclass[fleqn,usenatbib]{mnras}

\usepackage{graphicx}	
\usepackage{amsmath}	
\usepackage{amssymb}	

\title[GPA Applied to Galaxy Morphology ]{Gradient Pattern Analysis Applied to Galaxy Morphology}

\author[Rosa  et al.]{
R. R. Rosa$^{1}$\thanks{E-mail: rrrosa.inpe@gmail.com},
R. R. de Carvalho$^{2}$,
R. A. Sautter$^{1}$,
P. H. Barchi$^{1}$,
D. H. Stalder$^{1,3}$,
\newauthor
~T. C. Moura$^{2}$,
S. B. Rembold$^{4}$,
D. R. F. Morell$^{5}$,
N. C. Ferreira$^{1}$.
\\
$^{1}$Lab for Computing and Applied Mathematics, National Institute for Space Research (INPE), Brazil\\
$^{2}$Astrophysics Division, National Institute for Space Research (INPE), Brazil\\
$^{3}$NIDTEC/FP-UNA, San Lorenzo, Paraguay\\
$^{4}$Departamento de Fisica, CCNE, Universidade Federal de Santa Maria, Brazil\\
$^{5}$DCET, Universidade Estadual de Santa Cruz, Brazil
}

\date{Accepted XXX. Received YYY; in original form ZZZ}

\pubyear{2018}

\begin{document}
\label{firstpage}
\pagerange{\pageref{firstpage}--\pageref{lastpage}}
\maketitle

\begin{abstract}
Gradient pattern analysis (GPA) is a well-established technique for measuring gradient bilateral asymmetries of a square numerical lattice. This paper introduces an improved version of GPA designed for galaxy morphometry.  We show the performance of the new method on a selected sample of 54,896 objects from the SDSS-DR7 in common with Galaxy Zoo 1 catalog. The results suggest that the second gradient moment, $G_2$, has the potential to dramatically improve over more conventional morphometric parameters. It separates early from late type galaxies better ($\sim 90\%$) than the CAS system ($C\sim 79\%, A\sim 50\%, S\sim 43\%$) and a benchmark test shows that it is applicable to hundreds of thousands of galaxies using typical processing systems.

\end{abstract}



\section{Introduction}

Constraining almost 14 billion years of galaxy evolution from observations of galaxies as they are seen today is fraught with peril. One of the key aspects of any extragalactic investigation is the definition of an unbiased sample that includes reliable morphological types. Galaxy morphological properties result from not only the internal formation and evolution processes but also from interaction with the environment. Galaxies in groups or clusters may have diverse evolutionary paths compared to isolated ones, which is clearly reflected in their morphology. Therefore, classification of galaxies into a meaningful taxonomy system is of paramount importance for galaxy formation and evolution studies. 

Several attempts to objectively measure galaxy morphology have been tried. The most used system is based on  Concentration, Asymmetry, Smoothness, Gini and M20 (CASGM), presented in \cite{Abraham1994}, \cite{Conselice2000}, \cite{Conselice2003} and \cite{lotz2004}. The general rule for using a certain parameter to describe galaxy morphology is that it maximises the distinction between early and late type systems and minimize seeing effects. Such a parametrisation answers two immediate needs. First, to reproduce human classification by positioning the galaxies in the space of these parameters and second to establish a galaxy morphometry system that seek structures in the quantitative morphology parameter space that may yield clues on the physical reasons for the formation and evolution of galaxies. 

In the era of {\it Big Data} it is challenging to provide morphological information for a very large variety of galaxy types using only a few meaningful parameters \citep[e.g.][]{andrae2011}. The main goal of this paper is to present a new parameter, based on the gradient pattern analysis (GPA) formalism, that works complementary into the scope of galaxy morphometry.  GPA, since its appearance \citep{rosa1998,rosa1999}  and refinement \citep{rosa2003, costajunior2004}, became the main technique used for studying the gradient asymmetries of 2D spatio-temporal dynamics \citep{ramos2000}.  The method has been also applied for the characterization of pattern formation observed in porous silicon images obtained with scanning force microscopy \citep[see e.g.][]{daSilva2000}. Some of the applications were mainly concerned with structural analysis of patterns observed only in the spatial domain, which is also our main concern in this paper \citep[see e.g.][]{dasilvaSSC2000}. Regardless of the domain (dynamic or static), the central role in this technique is played by the so-called {\it gradient moments} \citep{rosa2003}. While there are well-established mathematical formulas for calculating the first gradient moment, $G_1$, \citep{rosa1999}, and the fourth, $G_4$, \citep{ramos2000}, there is no mathematical formalism in the literature for calculating the second and third gradient moments. In this work, we pay special attention to the definition and investigation of the second gradient moment, $G_2$, intended to be used in galaxy morphometry. We introduce this new parameter, based explicitly on vector norms, which improved the GPA by increasing: (i) its accuracy to distinguish an early from a late-type galaxy, and (ii) its computational performance when applied to large size images in a massively astronomical data set. 


\section{Data Used and Image Analysis}

Our sample is composed of galaxies from SDSS-DR7 (Sloan Digital Sky Survey - Data Release 7) in the redshift range  $0.03 < z < 0.1$, Petrosian magnitude in {\it r}-band brighter than 17.78 (spectroscopic magnitude limit), and $|b| \ge 30^{\circ}$, where $b$ is the galactic latitude. We restrict our sample to large systems - the area of the galaxy's petrosian ellipse is at least twenty times larger than the PSF area (r-band) for each corresponding object. This represents a sample of 57,841 galaxies in common with Galaxy Zoo 1 catalog \citep{zoo1}. We discarded, taking into account a field size of $5\times R_{P}$ \footnote{Petrosian radius}, 2,945 galaxy images due to several distinct problems: central double peak; galaxy at the edge of the field; many objects of similar brightness superimposed in the field; and merging galaxies. After pruning the sample from all these problematic images, we end up with 54,896 galaxies on which we focus our morphological study. According to Galaxy Zoo 1 catalog, 7,328 ($\sim$13\%) of these galaxies are ellipticals and 47,568 ($\sim$87\%) are spirals. We have used the weighted sample of galaxy classifications, which takes into account the fact that some users perform better when classifying galaxies (see \cite{zoo1} for more details). All image pre-processing is properly described in \cite{Barchi2018}. 

We developed a program called Cymorph written in Cython \footnote{C-Extensions for Python. The program is available under request} (\cite{Barchi2018} which measures: 1) Concentration (C) ; 2) Asymmetry (A); 3) Smoothness (S); 4) Entropy (H), is the Shannon entropy, namely the average amount of information resulting from a stochastic process - the clumpiest an image is the larger is its entropy\citep{ferrari2015}; and 5) GPA which will be described in the following Section. CyMorph measures C, A, and S following \cite{Conselice2000} and \cite {Conselice2003}.

\section{Gradient Pattern Analysis}

Considering that GPA has made only recently its entrance in astronomy, here we briefly describe its basics. The method was developed to estimate the local gradient properties of a set of points, which is generally represented in a 2-dimensional (2D) space. Astronomical images, represented by N $\times$ N pixels, can also be treated as a set of 3D vectors ${\bf m} = (x_i, y_i, I(x_i, y_i)),i = 1, . . . , N^2$, where $I(x_i, y_i)$ is the digital count measured on a given pixel, $x_i, y_i$. Representing images as count distributions, the local gradient is calculated as the first partial differences of  $I(x_i, y_i)$ with respect to each neighbour element in the lattice ${\bf m}$. The operation returns the $x$ and $y$ components of the two-dimensional numerical gradient, $\nabla M$, that can be characterized by each local vector norm and its orientation . The spacing between points in each direction is assumed to be 1 \citep{rosa1999, rosa2003}. In the GPA formalism, $\nabla M$ can be represented as a composition of the following four {\it gradient patterns} ({GP}): $\rm GP1$ (the lattice representation of the total vector distribution $\nabla M$); $\rm GP2$ (the lattice of the respective norms); $\rm GP3$ (the lattice of the respective phases); and $\rm GP4$ (the lattice of the respective complex numbers). For more details, see \cite{ramos2000} and \cite{rosa2003}. Indeed, for each type of lattice pattern from the set  $\lbrace \rm GP1, GP2, GP3, GP4 \rbrace$ we can calculate specific parameters which are defined by \citet{rosa2003} as the respective {\it gradient moments}: $\lbrace \rm {\it G_1, G_2, G_3, G_4} \rbrace$, where each is extracted from its respective lattice pattern, namely vector, norm, phase and complex representations.

In the GPA theory, there are already two consolidated measures for the first and fourth gradient moments. While the proposed first gradient moment, defined from a geometric approach by \citet{rosa1999}, works well to characterise both dynamic and static patterns, the proposed fourth gradient moment given by \citet{ramos2000} has been more effective only for dynamic patterns (that is, in the spatio-temporal domain). There are still no measures proposed in the literature for the second and third gradient moments. Thus, considering the main purpose of this work, we investigate how $G_1$ \citep{rosa2003} and $\rm GP2$ apply to galaxy morphological analysis.

\subsection{The First Gradient Moment}
To compute the first gradient moment, $G_1$, the method generally used is the {\it asymmetric gradient method} ($AGM$)\citep{rosa1998, rosa1999, rosa2003}. Once the $\rm GP1$, containing an amount of $V$ gradient vectors, has been established, the $AGM$ consists of the following four steps:

\begin{enumerate}

\item For a given tolerance for norm and phase of all vectors $V(i,j)$ of $\rm GP1$, we remove all symmetric pairs. The search for the pairs of vectors that have the same modulus and phase shifted by $\pi$, as shown in Figure 1, is concentric and starts from $V(1,1)$;

\item Once this operation is complete for all vectors, the $\nabla_{A} M(V_A)$ which contains an amount of $V_{A}$ asymmetric vectors is recorded. Note that, for a totally concentric symmetric pattern there will be no asymmetric vectors, that is $V_{A}=0$ (as the $\rm GP1$ of the symmetric example in Figure 1b);

\item  When $V_{A}>0$, we perform a Delaunay triangulation on $\nabla_{A} M(V_A)$, having the middle point of the asymmetric vectors as vertices;

\item Denoting $T_{A}$ as the resulting amount of edges of step 3, we can compute
\begin{equation}
G_1 = {{T_{A} - V_{A}} \over  {V_{A}}}.
\end{equation}

\end{enumerate}


\begin{figure}
	\centering 
	\includegraphics[width=.44\textwidth]{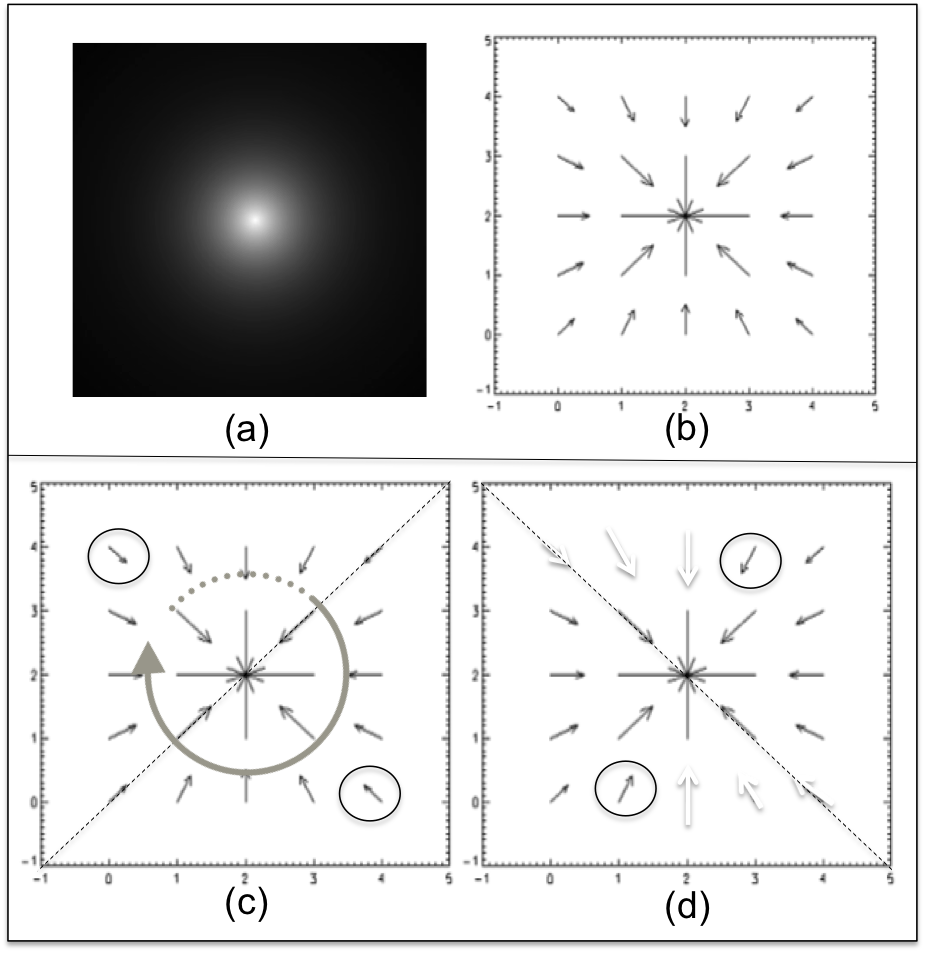}
    \caption{(a) a symmetric pattern generated from a 2D-gaussian function for a grid $6\times 6$; (b) the gradient pattern obtained from the operation $\nabla M$ on the image $M$, resulting $V=25$. The panels (c) and (d) show, as an example, two steps in removing the symmetrical vector pairs for elements V(1,1) and V(1,4), respectively. Note that, the search for any asymmetry takes as reference the four possible axes of bilateral symmetry: horizontal, vertical and the two diagonals.For this example, totally symmetric, $V_{A} = 0$.}
    \label{fig:example_figure}
\end{figure}

The operation that results in $V_A$ and $T_A$, for estimating $G_1$, is exemplified in Figure 2. This example considers a small matrix of size $5\times 5$, bilaterally asymmetric, where we show the sensitivity of $G_1$ to a variation by only one pixel, at the level of the standard deviation of all the values in the matrix. Figure 3 shows the same operation applied on two typical patterns, spiral and elliptical. Notice how dissimilar these two patterns are. The gradient moments were designed to properly quantify this dissimilarity.

The first gradient moment given by Equation 1, usually called the {\it gradient asymmetry coefficient},  has been used to characterize gradient patterns in reaction-diffusion processes  \citep{neto2000, rosa2000, assireu2002, costajunior2004}, molecular dynamics \citep{rosa2003} and stochastic surface growth \citep{baroni2006, rosa2007}. In all of these applications the dynamics of pattern formation is characterized by the temporal evolution of $G_1$ determined on each snapshot. 
When there is no bilateral asymmetries in the pattern, the total number of asymmetric vectors is zero, and then, by definition $G_1$ is null (as in the example of Fig.1). For a complex pattern composed by locally asymmetric structures we have $0<G_1<2$ defining different classes of irregular patterns and
for a random and totally disordered pattern,  $G_1$ has the highest value ($ G_1\rightarrow 2$ when $V_A\rightarrow \infty$). See details on the asymptotic behaviour of $G_1$ in \citet{rosa1999}. 

\begin{figure}
	\centering 
	\includegraphics[width=.46\textwidth]{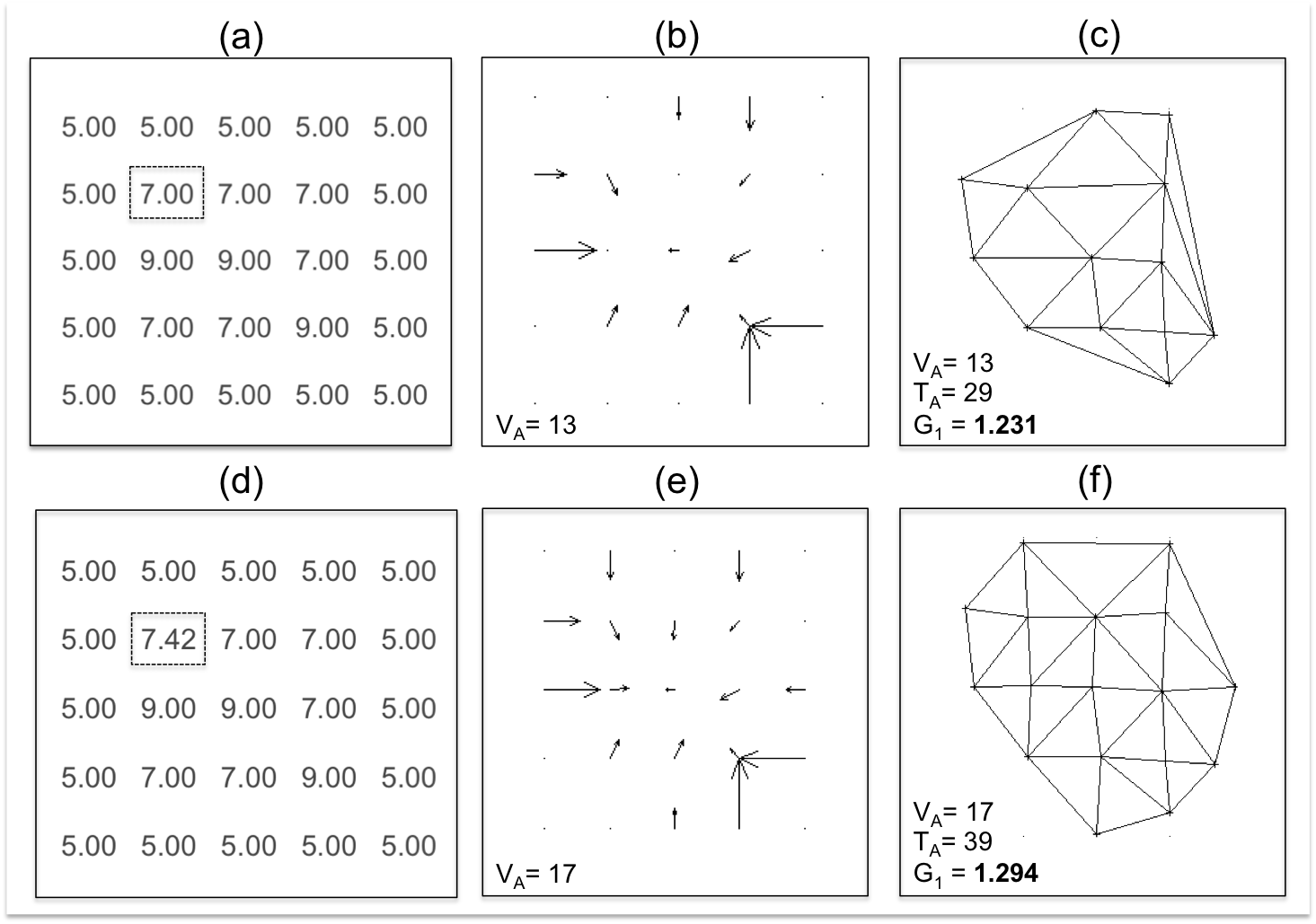}
    \caption{The application of GPA, for calculating $G_1$. (a) A specific matrix $M_{5\times5}$; (b) the respective gradient field of $M$ as computed by the original GPA; (c) the triangulation field and the values for $V_A$, $T_A$ and $G_1$ as given by Equation 1. (d)the same matrix, with a slight modification of the element $m_{2,2}$, where the standard deviation has been added to its original value. Panels (e) and (c) show the results after this pixel modification.}
    \label{fig:example_figure}
\end{figure}

More details on the role of the triangulation field in GPA are provided by \citep{rosa1999, rosa2003}. However, it should be noted from the example in Figure 2 that $T_A$ (or the triangulation field) is more sensitive to noise than $V_A$. This is due to the fact that fluctuations in the noise level affect the phases ($\rm GP3$) more than the norms ($\rm GP2$). Thus, for the applications in synthetic images from deterministic processes (for example, simulations of chaotic coupled maps and reactive-diffusive systems) the measurements of $G_1$ and $G_4$ are not hampered by the presence of noise. However, in the case of astronomical images, the performance of these two parameters, including a possible parameter $G_3$, which are explicitly based on the phase values ($\rm GP1$, $\rm GP3$ and $\rm GP4$), will be affected by the noise. Therefore, an alternative to apply GPA to galaxy morphology is to seek an unprecedented measure for $\rm GP2$. 

\begin{figure}
	\centering 
	\includegraphics[width=.42\textwidth]{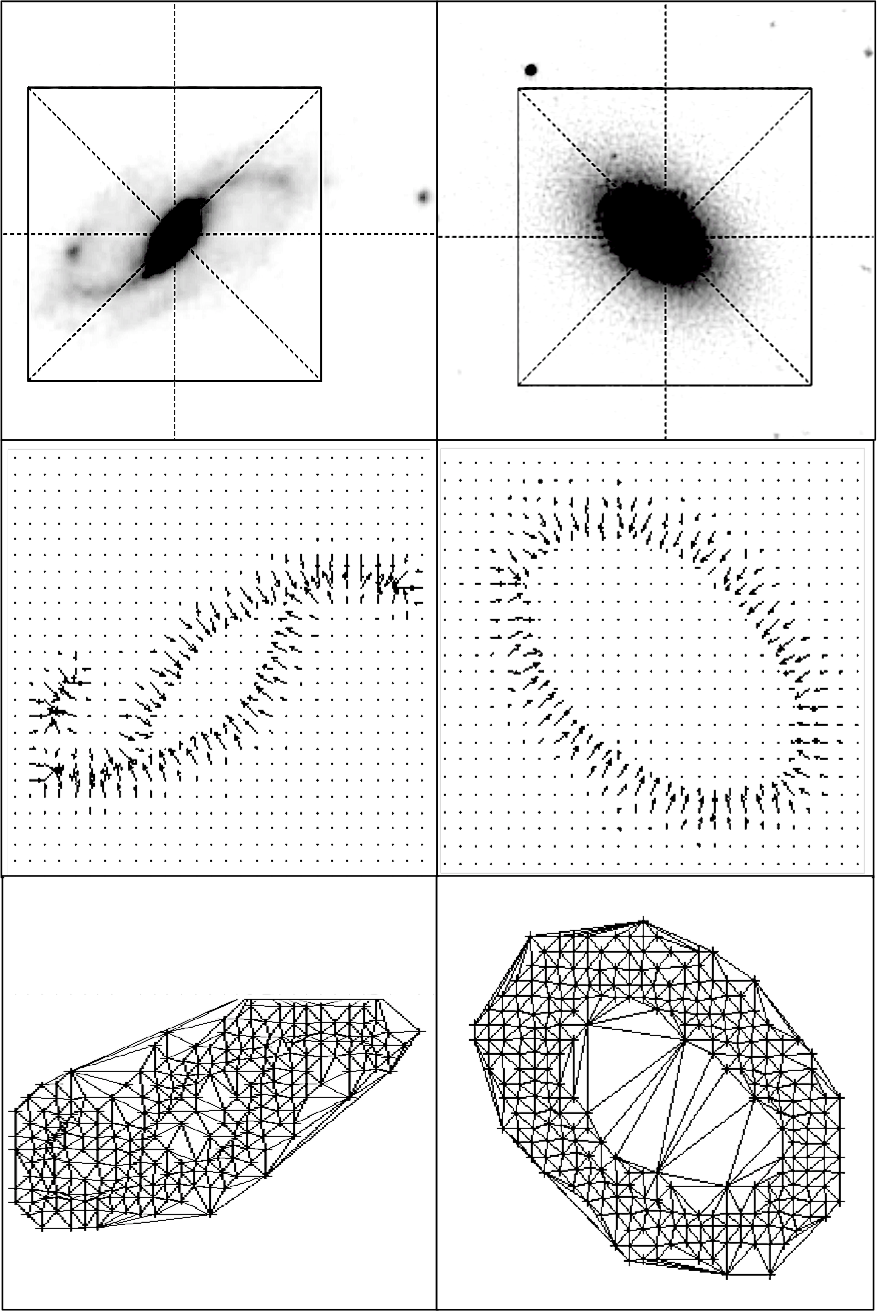}
    \caption{The application of GPA, for calculating $G_1$, on two galaxies selected from our {\bf SDSS DR-7, r-band, sample}  - a spiral on the left and an elliptical on the right. The respective gradients containing $V_A$ vectors are shown in the intermediate panels and the triangulation fields, containing $T_A$ edges, are shown in the bottom panels. Images are in reverse color contrast for better viewing.}
    \label{fig:example_figure}
\end{figure}


\subsection{The Second Gradient Moment}

Considering the difficulties inherent to the morphology of galaxies through digital image analysis, such as the need of segmentation, some improvements were incorporated into the first GPA operation that deals with the generation of the gradient field from the image: (i) the gradient should be calculated disregarding the border elements, thus generating a gradient field with $(N-2)\times  (N-2)$ positions when the matrix is square, and  (ii) allow the calculation on a rectangular matrix. In addition, an appropriate GPA measure for astronomical images should be invariant to image size and less sensitive to noise than $G_1$. Taking into account such conditions we derive an efficient measure based only on the lattice of the norms $\rm GP2$. Thus, the second gradient moment within the GPA formalism is introduced here as follows:
\begin{equation}
G_{2} = \frac{V_A}{V}\left (2-\frac{|\sum_{i}^{V_A} v_i|}{\sum_{i}^{V_A} |v_i|} \right )
\label{gal}
\end{equation}

\noindent where $V$ is the total amount of gradient vectors and $V_A$ is the amount of asymmetric vectors after removing all the symmetric pairs, representing the same quantities previously defined for the determination of $G_1$. Then, the $\sum_{i}^{V_a} v_i$ is the asymmetrical vector sum and $|v_i|$ is the $i^{th}$ asymmetrical vector norm. Notice that for misaligned vectors, the vectorial sum tends to zero. More formally, we can write $|\sum_{i}^{V_A} v_i| = 0$, then according to equation 1, $G_{2} = 2\frac{V_A}{V}$. Whereas if $K$ vectors with same moduli are aligned, $|\sum_{i}^{V_A} v_i| = K|v_i|$, and $\sum_{i}^{V_A} |v_i| = K|v_i|$. Therefore, $G_{2}= \frac{V_A}{V}\left ( 2-\frac{K|v_i|}{K|v_i|} \right ) = \frac{V_A}{V}$. This means that this operator considers the proportion of asymmetrical vectors, and also, without using explicitly the phases ($GP_3$), the correspondent  alignment rate. Higher $G_2$ values mean that the gradient lattice has many misaligned asymmetrical vectors and then a high diversity for the values in the lattice $\rm GP2$. 
Therefore, the calculation of $G_2$ on canonical matrices demonstrates its ability to characterise the basic asymmetry conjectures investigated in \cite{rosa1999} using $G_1$, however with less sensitivity to the noise due to phase fluctuations imprinted in the triangulation field.  For the example in Figure 2, the result for $G_2=0.656$ is invariant considering both matrices.

The operation for computing $G_2$, via Eq. 2,  presents the following improvements compared to  $G_1$: ({\it i}) For the same type of gradient pattern, the value of $G_2$ is invariant to the size of the matrix; ({\it ii}) More appropriately, it does not consider the elements of the matrix border for calculating the gradient; ({\it iii}) Can be applied, without loss of generality, to rectangular matrices; and ({\it iv}) It is less sensitive to noise and faster because it avoids triangulation.

\section{Galaxy Morphometry using GPA}

The morphological analysis of the 54,896 objects, as classified by Galaxy Zoo 1, has been done on the basis of the parameters $G_1$, $G_2$ and $H$ defining our new system, and $C$, $A$ and $S$ which defines the commonly used CAS system \citep[]{Conselice2000}.  The respective histograms are shown in Figure 3 where the red (blue) line refers to elliptical(spiral) galaxies. 

We present a comparative analysis between well established parameters, C, A, S, used in several galaxy morphology studies \citep[e.g.][]{Conselice2000, ferrari2015},
and those proposed in this investigation, $G_1$, $G_2$ and $H$. A given parameter is considered a useful morphological indicator when it separates early and late type galaxies the best possible. Therefore, it is of paramount importance to objectively define separation. 
In our case, we estimate how far apart two histograms are (see Figure 4), using the index $\delta_{GHS}$ which is calculated from the GHS (Geometric Histogram Separation) algorithm \citep{box2017}. This algorithm determines separation using only the geometric characteristics of a binomial proportion represented by histograms. The GHS input parameters are: $A_B$ (blue histogram area), $A_R$ (red histogram area), $A_{BR}$ (intersection area between $A_B$ and $A_R$), and the respective heights for $A_B$, $A_R$ and $A_{BR}$: $h_B$, $h_R$, and $h_{BR}$. The separation is then defined as:
\begin{equation}
\delta_{GHS} = {{(1 - {A_{BR} \over A_{B}+A_{R}+A_{BR}})^{1/2} + ({{h_a + h_b - 2h_c}\over{h_a+h_b}})} \over 2}.
\end{equation} 

\begin{figure}
\centering
\includegraphics[width=.46\textwidth]{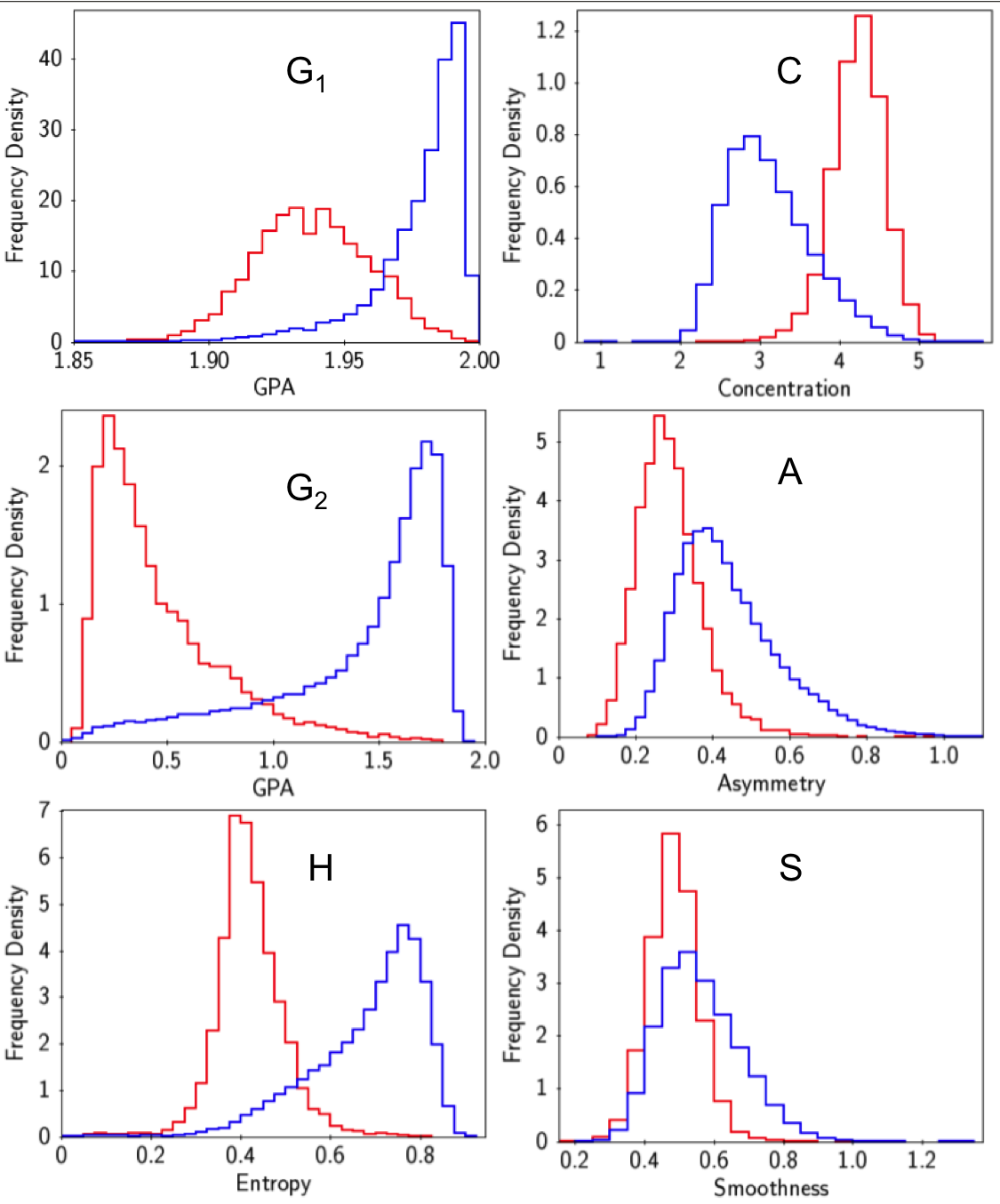}
\label{preprocessingPip}
\caption{Histograms for the six morphological parameters used to classify galaxies. {\bf Elliptical (red) and Spiral (blue).}}
\end{figure}

\begin{table}
\centering

\caption{The morphological parameters performance $P=\delta_{GHS}/\langle t\rangle$. The $\delta_{GHS}$ is calculated applying equation 3 to the histograms of figure 4 and $\langle t\rangle$ is the average processing time, in seconds, for each parameter. To highlight the improvement in the new GPA approach introduced from Equation 2, we set the time values for $G_1$ and $G_2$ in bold. \label{tab1}}

\begin{tabular}{llll}
\hline

\textbf{Parameter}  & \textbf{$\delta_{GHS} (\%)$} & \textbf{$\langle t\rangle$ (s)} 
& \textbf{$P (s^{-1})$}\\
\hline


{\it GPA} (${\rm \bf  G_{2}}$)   & 90.4   & {\bf 1.840} & 49.1 \\
{\it Entropy} ($\rm H$)         & 86.9   & 2.035 & 42.7 \\
{\it GPA} (${\rm \bf G_{1}}$)   & 82.3   & {\bf 1.915} & 42.9 \\
\hline
{\it Concentration} ($\rm C$)   & 78.9   & 2.262 & 34.9\\
{\it Asymmetry} ($\rm A$)     & 49.6   & 1.854 & 26.8\\
{\it Smoothness} ($\rm S$)    & 42.7   & 1.852 &  23.1\\

\hline 
\end{tabular}
\end{table}

The respective values, presented here as percentages, are listed in Table 1, where  0\% indicates complete superposition and $100\%$ indicates total separation. The results in Table 1 attest that the parameter $G_2$ is not only more effective than $G_1$ but also works far better than the parameters of the CAS traditional system and slightly better than H. It is worth mentioning the processing time, since we plan on using CyMorph extensively on data already available, like SDSS \citep{2017arXiv170709322A}, KiDS (Kilo Degree Survey) \citep{dejong2015}, etc. The average processing time per galaxy on the entire sample, $\langle t\rangle$ in seconds, is presented for each parameter. This computation was performed on a machine based on Intel i8, 3.5 GHz, RAM  62GB and HD 2TB. The parameter performance, $P(s^{-1})$, has been defined as $\delta_{GHS}/\langle t\rangle$. This result is critically important, considering the wealth data made available by ongoing and upcoming surveys. If we compare the best parameter of our system, $G_2$, with the best of the CAS system, C, we conclude that in order to process one million galaxies it would take $\sim$22 days to estimate $G_2$ as opposed to $\sim$26 days to estimate C. This difference is very significant considering that the new upcoming surveys will gather hundreds of millions of galaxies, depending on the limiting magnitude used. Also, the difference in $\delta_{GHS}$, $\sim$12\%, considering $G_2$ and C, is very significant when the main purpose is to have robust morphology for a large sample.




\section{Discussion}

Morphology is a key ingredient in the process of selecting a sample of galaxies for studying the physical mechanisms responsible for shaping the galaxies as we observe today. Also, considering that the following decades will be dominated by photometric (image) rather than spectroscopic data (e.g. LSST, Pan-STARRS, etc.), it is critical to have robust measurements that capture the essential morphological information and avoid redundancy.

In this Letter, we present a new morphological parameter based on the gradient pattern analysis formalism.
We examine a sample of bright galaxies from SDSS-DR7 in the redshift domain of $0.03 < z < 0.1$ and for which the area is twenty times larger than the PSF's. These criteria resulted in a sample of 54,896 systems in common with Galaxy Zoo, which is used here as the true morphology provider. From a comparative analysis considering the application of other five parameters we find that $G_2$ is the one providing the largest separation between early and late type galaxies ($\sim90\%$). Due to the achieved success, two points should be highlighted: (i) this new parameter, based on a measurement on the gradient of the image, is not equivalent to any other used in the literature. The measurement, unlike the ones that operate directly on the image counts, retains the signature of the asymmetric fluctuation among the gradient vectors; (ii) The benchmark presented here shows how feasible it is to apply the new parameter to classify millions of galaxies in an automatic fashion using currently available computational resources.
We are currently investigating the application of supervised and unsupervised machine learning clustering methods, using the GPA, $H$ and an improved version of $CAS$\citep{Barchi2018}, to provide unbiased morphological classification for hundreds of millions of galaxies.


\section*{Acknowledgements}
{\small The authors are grateful to M. Soares-Santos for the critical reading of the manuscript. R.R.dC. and R.R.R. acknowledge financial support from FAPESP through grant \# 2014/11156-4.}




\bibliographystyle{mnras}
\bibliography{references} 
\bsp	
\label{lastpage}
\end{document}